\pdfoutput=1
\documentclass[12pt]{iopart}
\usepackage{iopams}
\usepackage[latin1]{inputenc}
\usepackage{graphicx}
\usepackage[hypertex]{hyperref}

\newcommand{\average}[1]{\left<{#1}\right>}
\newcommand{\C}{\mathrm{c}}
\newcommand{\E}{\mathrm{e}}
\newcommand{\leader}{{\mathrm{leader}}}
\newcommand{\given}{\mathrel{|}}
\begin{document}

\title[Microevolution in a changing environment]{Modeling microevolution in a changing environment: The evolving quasispecies and the Diluted Champion Process}

\author{Ginestra Bianconi\dag, Davide Fichera\ddag\, Silvio Franz\ddag, and
Luca Peliti\S\footnote[4]{To whom correspondence should be
addressed.}}

\address{\dag\ Physics Department, Northeastern University, Boston MA 02115 USA}

\address{\ddag\ Laboratoire de Physique Théorique et Modèles Statistiques,  CNRS and~Université~Paris-Sud, UMR~8626, Bât.~100, 91405~Orsay~cedex France}

\address{\S\ Dipartimento di Scienze Fisiche and Sezione INFN, Università ``Federico II'',
Complesso Monte S. Angelo, 80126 Napoli Italy}

\begin{abstract}
Several pathogens use evolvability as a survival strategy against
acquired immunity of the host. 
Despite their high variability in time, some of them exhibit quite low
variability within the population at any given time, a somehow
paradoxical behavior often called the evolving quasispecies. 
In this paper we introduce a simplified model of an evolving viral
population in which the effects of the acquired immunity of the host
are represented by the decrease of the fitness of the corresponding
viral strains, depending on the frequency of the strain in the viral
population. The model exhibits evolving quasispecies behavior in a
certain range of its parameters, and suggests how punctuated evolution
can be induced by a simple feedback mechanism.
\end{abstract}


\ead{peliti@na.infn.it}

\maketitle

\section{Introduction}
Pathogenic viruses, in order to survive and successfully reproduce
have to fight  
against the immune system of their host organisms. Some viruses use
evolvability as a successful strategy to escape acquired immunity. In the
presence of adaptive response in the host newly arising mutants  can acquire a
competitive advantage with respect  
to the wild type. Neverthless, in some viral populations one often  
observes ``quasispecies'' behavior, in which individuals are strongly
similar to one another.  

A prototypical example of this behavior is exhibited by the 
Influenza A virus, which makes use of high antigenical 
variability (genetic drift) to escape acquired immune response.  
Virus strains circulating in different epidemic seasons, in spite of eliciting a
certain amount of cross-response by the immune system of the hosts, mutate 
fast enough to be able to infect the same host several times in
the course of its life.  However in any given observed epidemics, the
viral population is sufficiently concentrated around a well defined
strain that effective vaccines can be prepared. Moreover, antigenic
changes are punctuated: antigenic assays show that the sequences of
the dominant circulating type H3N2 fall into temporally correlated
clusters with similar antigenic properties~(\cite{Smith}). This
behavior, which has been described as an evolving viral quasispecies,
contrasts with the prediction of naive models of viral evolution (see,
e.g., \cite{Girvan}),
where the interaction with the immune system leads to proliferation of
strains with different antigenic properties and, consequently, to the
impossibility of preparing effective vaccines.  It has recently been experimentally shown~\cite{bet-hedging} that some bacteria in a changing environment can develop a genotype which produces highly variable phenotypes, presumably adapted to different environments. One may speculate whether such a bet-hedging strategy is available to viruses, due to the small size of their genetic material and their high mutation rate, especially for RNA-based viruses.

A different example is the one of the in-host evolution of the HIV,
which exploits high mutation rates to counter the immune adaptation of its host
(see, e.g., \cite{Shankarappa96,Shankarappa99}). Despite its high
mutation and substitution rates, the viral population shows low levels
of differentiation during most of the asymptomatic phase.  

Several solutions have been proposed to this evolutionary
puzzle, based on the analysis of mathematical models.
In~\cite{Ferguson}, elaborating on previous theory~(\cite{Webster}),
it was proposed that in the case of influenza A, a short-time
strain-trascending immunity could give account for the quasispecies
behavior. It was observed in~\cite{Tria} that this short-term immunity
would be ineffective in the absence of heterogeneities in
infections, due either to the structure of network of infective
contacts or to variations in the
infective efficacy of different strains.  An alternative explanation
was put forward in~\cite{Koelle} based on the analysis of a model,
where the heterogeneity of the neutral phenotypic network of the evolving virus
is taken into account. An important feature in influenza epidemics
seems to be the directionality of its propagation. According to the
analysis of~\cite{Rambaut} and \cite{Russell} yearly epidemics typically start from
seeds in 
South-East Asia before spreading worldwide.  This suggests the
presence of strong bottlenecks in the viral population that could lead
to relatively small effective population sizes and to a strong
reduction of genetic variability.

Models can suggest mechanisms that can produce the observed
evolutionary pattern. However, even the simplest individual based
model of viral evolution in a host population (\cite{Tria}) is too
complicated to lead to a detailed exploration of parameter space.  It
is not clear which conditions on evolutionary and
epidemiological parameters allow for the evolving quasispecies
behavior. In particular we will be interested to the dependence on the
population size. We wish to ascertain whether in the presence of the
infectivity reduction due to the immune response of the host,  the
quasispecies behavior can appear 
for some parameter ranges in the infinite
population, or is only possible in sufficiently small populations.
This last possibility would emphasize the importance of population
bottlenecks in the ecology of the virus. 

In this paper we introduce a simplified model of evolving viral
population that keeps into account the interaction between virus and
host immunity in an effective way. The effectiveness of reproduction of
a given viral type at a given time depends on its age and its past
frequency in the host population. 
We get a model that is simple enough to allow to study at least numerically,
the dependence on the different parameters, and in 
particular on the population size. Our findings suggest that 
quasispecies behavior and punctuated equilibria are only possible for
small enough populations, whereas, if the model parameters are not
scaled with the population size, one always has proliferation in the
large population limit. 

The organization of the paper is the following: in
section~\ref{sect:TheModel} we define our model as an extension of
Kingman's house of cards model, where strains are represented by
self-reproducing entities and the immune memory of the host is taken
into account by the decrease of the strain fitness with time. 
In this model, the fitness of a mutant is independent of that of
its parental strain. This assumption is justified in our approach
by the fact that since the parental fitness decreases very quickly due to the immune adaptation of the host, we do not expect long-lived fitness
correlations to exist among related strains.
In section~\ref{h0} we recall some known results about the dynamics of
the model in the absence of memory.  In section \ref{fitdec} we study the
effect of the immune memory in numerical simulations and show the
existence of the evolving quasispecies regime. In~\ref{dyn} we analyze in more detail the behavior of the system in the evolving quasispecies regime. In \ref{dcp} we
introduce a simple effective stochastic process that helps in
understanding the dynamical behavior of the quasispecies
regime. Finally some conclusions are drawn.

\section{The Model}
\label{sect:TheModel}
We consider here a model of an evolving viral population of constant
size consisting of $N$ individuals. The population evolves according
to a Wright-Fisher process for asexually reproducing 
populations~(\cite{Wright,Fisher}). At each discrete time step $t$ the 
population reproduces and is completely replaced by its progeny. Each
individual is characterized by its strain label $S$.
The expected offspring size of an individual belonging to strain $S$
is proportional 
to the Wrightian fitness $w_S(t)$ of its strain. The fitness
$w_S(t)$ of a strain $S$ depends on its intrinsic fitness $w_S^0$,
proportional to its 
basic reproductive number, and changes with time as described below.
In what follows we shall also use the notation
$w_S=\rme^{f_S}$ and call $f_S$  the Fisher fitness of $S$.
As it is well known~(\cite{DerridaPeliti}) the Wright-Fisher
model can be described as a process where 
each individual $j$ at generation $t+1$
``chooses'' its parent $i$ at generation $t$ with probability equal to
$w_{S_i}(t)/\sum_{j=1}^N w_{S_j}(t)$.  Consequently, for large $N$, the
offspring size of an individual $i$ is a random Poissonian variable
with average  $w_{S_i}(t)/\average{w}_t$.

At each reproduction event, a mutation can take place with probability
$\mu$. Upon mutation, a new strain $S'$ appears, and its intrinsic
fitness $w_{S'}^0$ is drawn from a fixed probability distribution
$\rho(w)$~(\cite{Tria,Kingman78,FranzPeliti,Shraiman}), independently
for each mutation event: this corresponds to
the infinite allele approximation where no back mutations are possible.
The fitness $w_{S'}$ of the newly formed strain is equal to its
intrinsic fitness $w_{S'}^0$. If fitnesses did not depend on time, the
model would coincide with Kingman's house-of-cards model~(\cite{Kingman78}).

The effect of acquired immunity of the host population on the viral
reproduction  is represented by letting the fitness $w_S$ of each
strain $S$ decrease at each generation with a rate proportional to the
number $N_S(t)=N n_S(t)$ of individuals belonging to that strain in
the population. For simplicity we consider an exponential decrease of
the fitness: 
\begin{eqnarray}
\label{eqn:evolv_fitness}
w_S(t+1)=w_S( t )\, \rme^{-h N_S(t)/N},
\label{uno}
\end{eqnarray}
where $h$ is a parameter which determines the rate of fitness decrease. 

In the following we will describe mainly the case in which $\rho(w)$
is a log-normal distribution
$\rho(w)\propto \rme^{-\log^2 w/2a^2}/w$. We have also investigated
the model with the uniform fitness distribution finding the same
qualitative results. 

\section{Dynamic behavior in the absence of immunity}
\label{h0} 
We start by reviewing the behavior of the model for $h=0$, i.e., in
the absence of immune memory of the host~(\cite{Kingman78,Shraiman,Krug}).
In this case, in the limit of large populations $N\to\infty$,
it is possible to derive
an exact equation for the evolution of the fraction $x_t(w)$ of
individuals with fitness $w$ at time $t$, namely 
\begin{eqnarray}
x_{t+1}(w)=\frac{w}{\average{w}_t}\,(1-\mu)x_t(w)+\mu\,\rho(w),
\label{mf}
\end{eqnarray}
where $\average{w}_t=\int \rmd w\; w x_t(w)$. The nature of the
solution of this equation depends on the properties of $\rho(w)$. If
$\rho(w)$ has a compact support (i.e., $\rho(w)=0$ for $w>w_{\max}$),
the equation admits a stationary state satisfying  
\begin{eqnarray}
x(w)=\frac{\mu\,\rho(w)}{1-(1-\mu)\,w/\average{w}_t}. 
\end{eqnarray}
This distribution is analogous to the Bose distribution, with $w$
playing the role of the Boltzmann factor. Therefore, if $\int \rmd w\;
\rho(w)/\left(1-w/w_{\max}\right)<\infty$, the Einstein
condensation can take place. This transition is known as the error threshold
in evolutionary models, and separates a poorly adapted phase
at high mutation rate from a well adapted phase at low mutation rate.
In the well adapted (condensed) phase, found for $\mu<\mu_\C$, a
finite fraction $\nu$ 
of the population has the maximum fitness $w_{\max}$. Here $\nu$ and
$\mu_\C$ are given by
\begin{eqnarray}\label{muc:eq}
\nu=1-\frac{\mu}{\mu_\C};\qquad
\mu_\C^{-1}=\int_0^{w_{\max}}\rmd w\; \frac{\rho(w)}{1-w/w_{\max}}.
\end{eqnarray}
In this phase the reproduction rate of the individuals with 
maximal fitness equals one: $(1-\mu)w_{\max}/\average{w}=1$, while 
all the others have reproduction rates smaller than one. 
Conversely, in the poorly adapted (uncondensed) phase, less than
maximally fit individuals can reproduce.  
If the integral $\int \rmd w\; \rho(w)/\left(1-w/w_{\max}\right)$
is divergent no error threshold appears.  

\begin{figure}[ht]
\begin{center}
\includegraphics[width= 0.7 \textwidth]{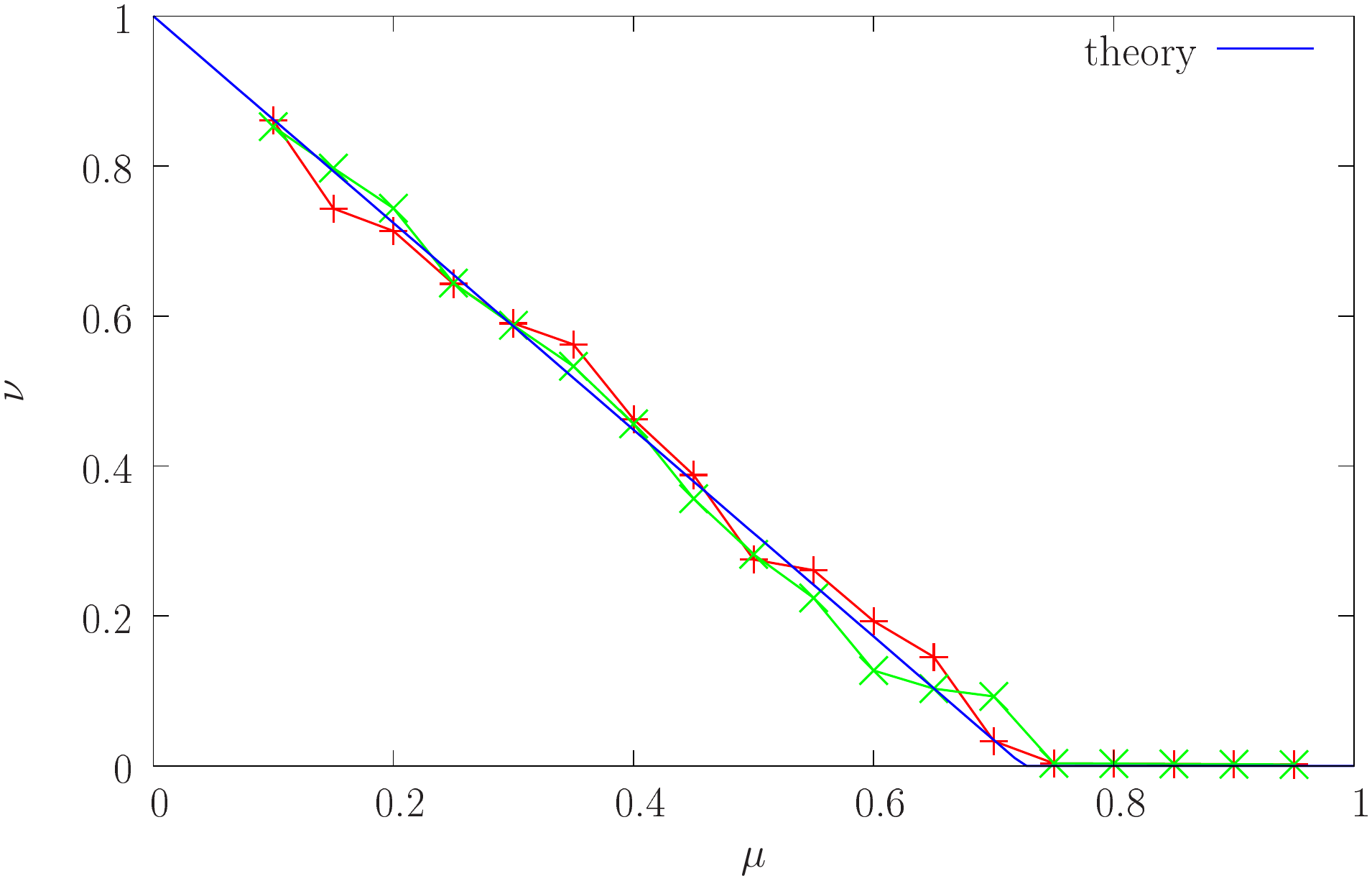}
\caption{Condensate fraction $\nu$ as a function of the mutation rate $\mu$ for
  two independent populations of 1000 individuals evolving for $t=5000$
  generations, together with the theoretical result. The data display a
  well defined error threshold in agreement with the theory.  } 
\label{fig1}
\end{center}
\end{figure}

The situation where the fitness distribution $\rho(w)$
extends to infinity is more complex and interesting. In this case,
equation~(\ref{mf}) does not admit a 
stationary solution and one obtains instead a run-away. However,
in contrast with the case of compact support, the
dynamics for a finite population with $N\gg 1$ individuals can now be
substantially different from the infinite population limit. In fact,
the evolution process can be described as a non stationary record
process where condensates of a given fitness are replaced by ones
of higher fitness~(\cite{Krug}), with persistence times that becomes
longer and longer as the fitness increases. It has been noticed in~\cite{Shraiman} that, however, an uncondensed phase can be metastable for long
times. Let us denote by $w_{\mathrm{mut}}(t)$ the maximum fitness appearing among the mutants in $t$ generations. As at each time step a number $N\mu$ of new mutants appear,
the maximum fitness $w_{\mathrm{mut}}(1)$ for one generation satisfies
\begin{eqnarray}
N\mu\int_{w_{\mathrm{mut}}(1)}^\infty \rmd w\; \rho(w)\simeq 1.
\end{eqnarray}
If the reproduction rate $(1-\mu)w_{\mathrm{mut}}(1)/\average{w}$
is smaller than one, the uncondensed phase will be stable. This quantity depends on $t$ via the average $\average{w}$. Then the
stability time can be determined by the condition that the maximal fitness
of mutants over an interval of $t$ generations, as determined by the
relation $N\mu t\int_{w_{\mathrm{mut}}(t)}^\infty \rmd w\;
\rho(w)\simeq 1$ has   
a reproduction rate equal to one. In this way one gets an effective
$N$- and $t$- dependent error threshold, which can be approximately
described by equations
\begin{eqnarray}
\nu=1-\frac{\mu}{\mu_\C};\qquad
\mu_\C^{-1}=\int_0^{w_{\mathrm{mut}}(1)}\rmd w\; \frac{\rho(w)}{1-w/w_{\mathrm{mut}}(t)}.
\label{p1}
\end{eqnarray}
In figure~\ref{fig1} we compare this analysis with numerical
simulations for the lognormal distribution with parameter $a=0.3$,
showing the fraction $\nu$ of the condensate for two independent populations of
$N=1000$ individuals evolving for $t=5000$ generations, together with
the predictions of equation~(\ref{p1}) as a function of the mutation
rate.

\section{Introducing the fitness decay} 
\label{fitdec}
\begin{figure}[ht]
\begin{center}
\includegraphics[width= 0.7 \textwidth]{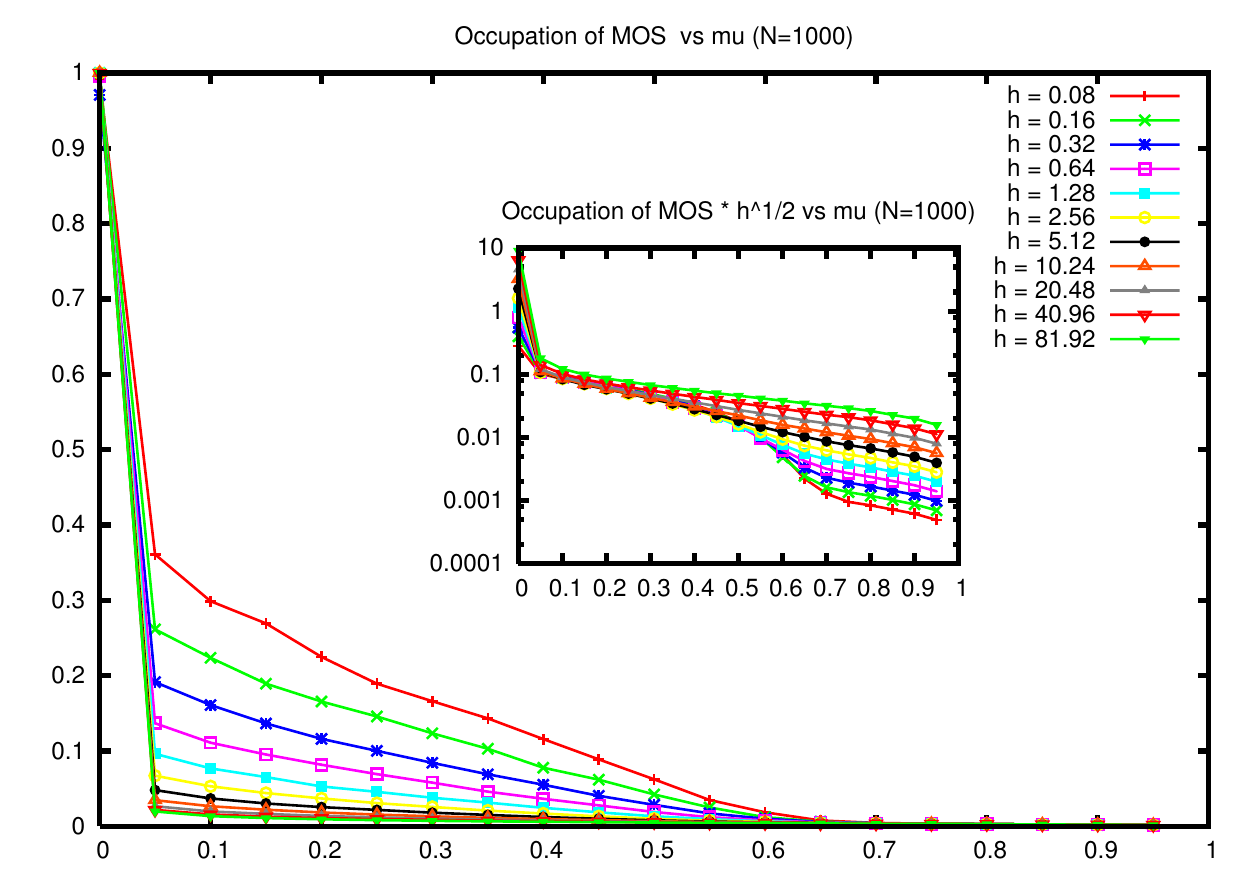}
\caption{
Occupation fraction $\nu$ of the leader for $N=1000$,
$a=0.3$ and several values of $h$, as a function of $\mu$. The error
threshold takes place at an $h$-independent value of $\mu$ if $h$ is
not too large.  
In the inset we plot $\nu h^{1/2}$ exhibiting the proportionality of
$\nu$ to $h^{-1/2}$ in the condensed phase. 
 }
\label{fig2}
\end{center}
\end{figure}

\begin{figure}[ht]
\begin{center}
\includegraphics[width= 0.7 \textwidth]{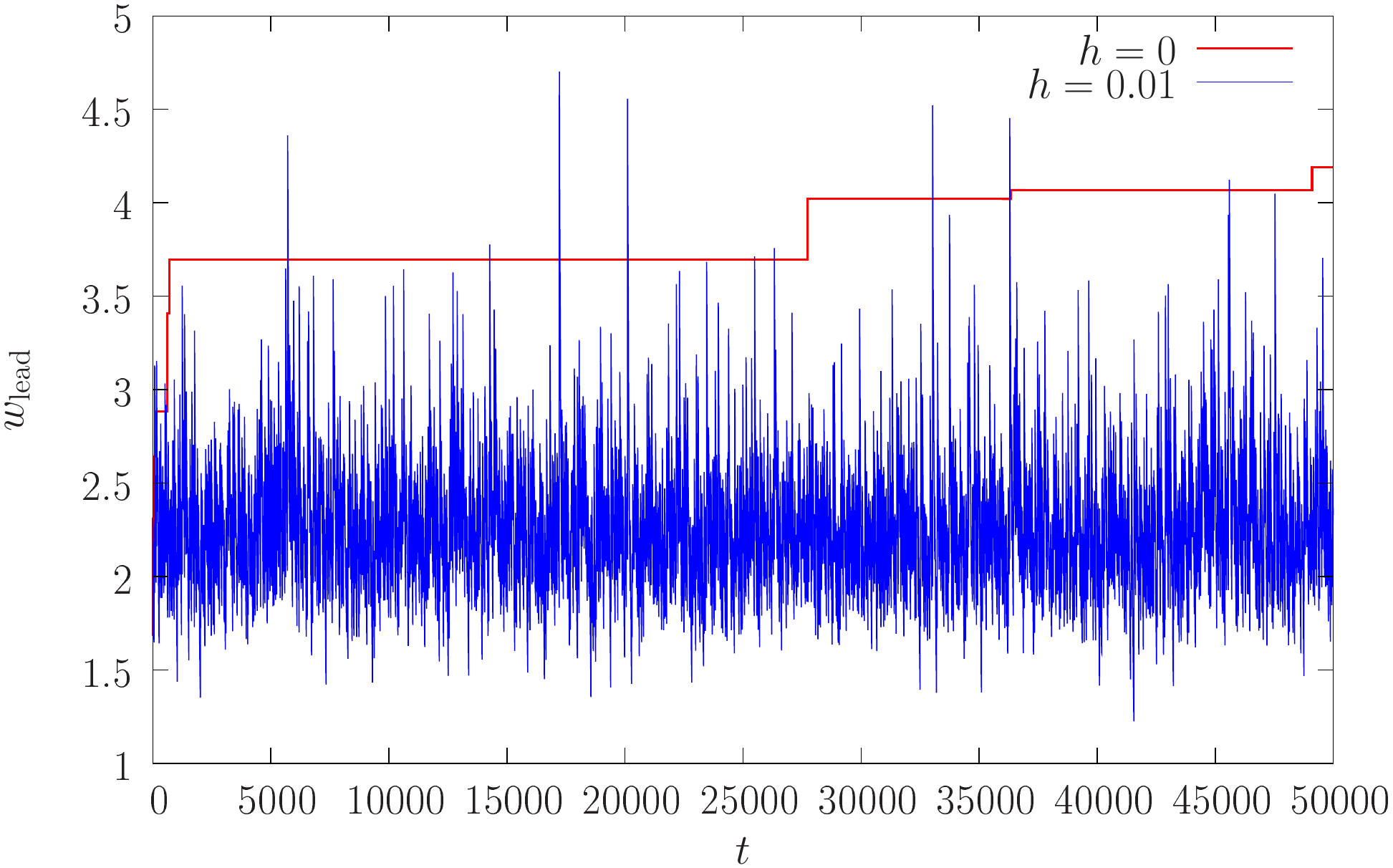}
\caption{Fitness $w_{\mathrm{lead}}$ of the leader as a function of
  time for the lognormal fitness distribution with $a=0.3$, $\mu=0.1$,
  and for $h=0$ and $h=0.01$ respectively. For $h=0$ the
  fitness follows a record process increasing on average with time. As
  soon as $h>0$ the process becomes stationary.} 
\label{fig3}
\end{center}
\end{figure}

\begin{figure}[ht]
\begin{center}
\includegraphics[width= 0.7 \textwidth]{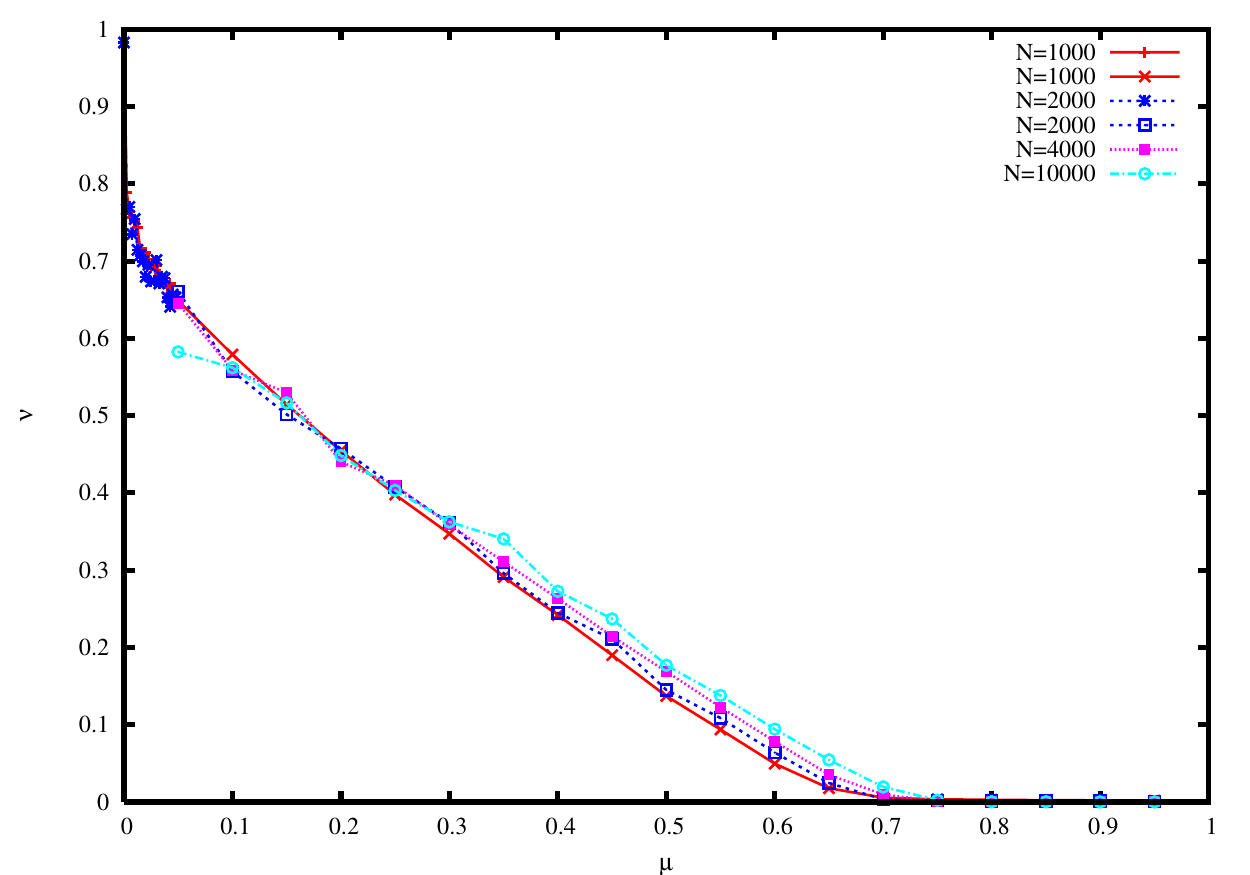}
\caption{Occupation $\nu$ of the leader for $h=0.01$ as a function of $\mu$
  for $N=1000,2000,4000,10000$. The error threshold slowly moves to
  higher values of $\mu$ as $N$ increases.} 
\label{fig4}
\end{center}
\end{figure}

\begin{figure}[htb]
\begin{center}
\includegraphics[width= 0.7 \textwidth]{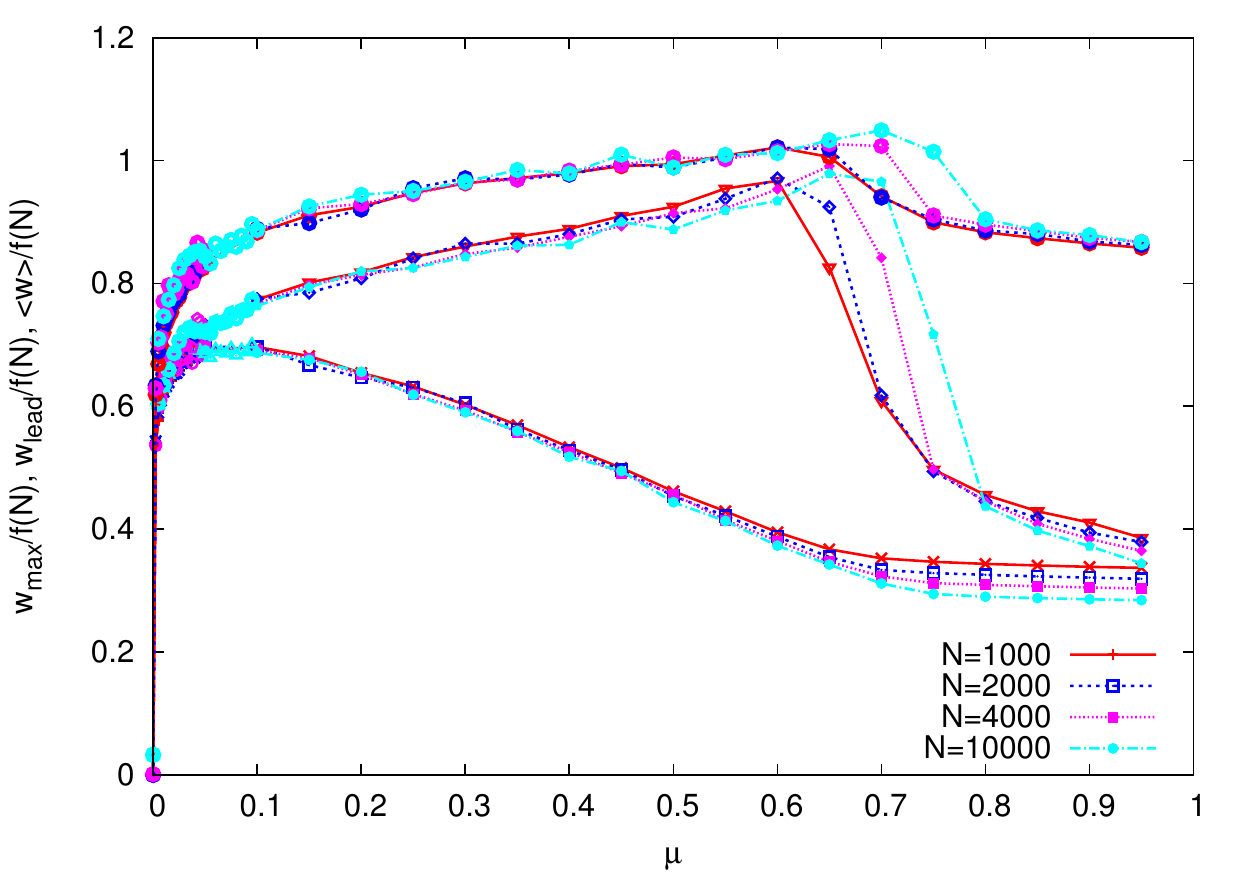}
\caption{%
Maximal $w_{\mathrm{max}}$ (upper line), leader $w_{\mathrm{lead}}$ (middle line) and   average $\average{w}$ (lower line) fitness 	for $h=0.01$, as a function  of $\mu$, for $N=1000,2000,4000,10000$. The fitnesses are rescaled by the factor $f(N)=\exp\left(\sqrt{2a^{2}\log(N/\sqrt{2\pi a^{2}})}\right)$, where $a=0.3$ is the width of the distribution of $\log w_{0}$, expected on the basis of extreme-value statistics. The error threshold is slowly displaced to higher values of $\mu$ as $N$ increases.}
\label{fig5}
\end{center}
\end{figure}

\begin{figure}[ht]
\begin{center}
\includegraphics[width= 0.7 \textwidth]{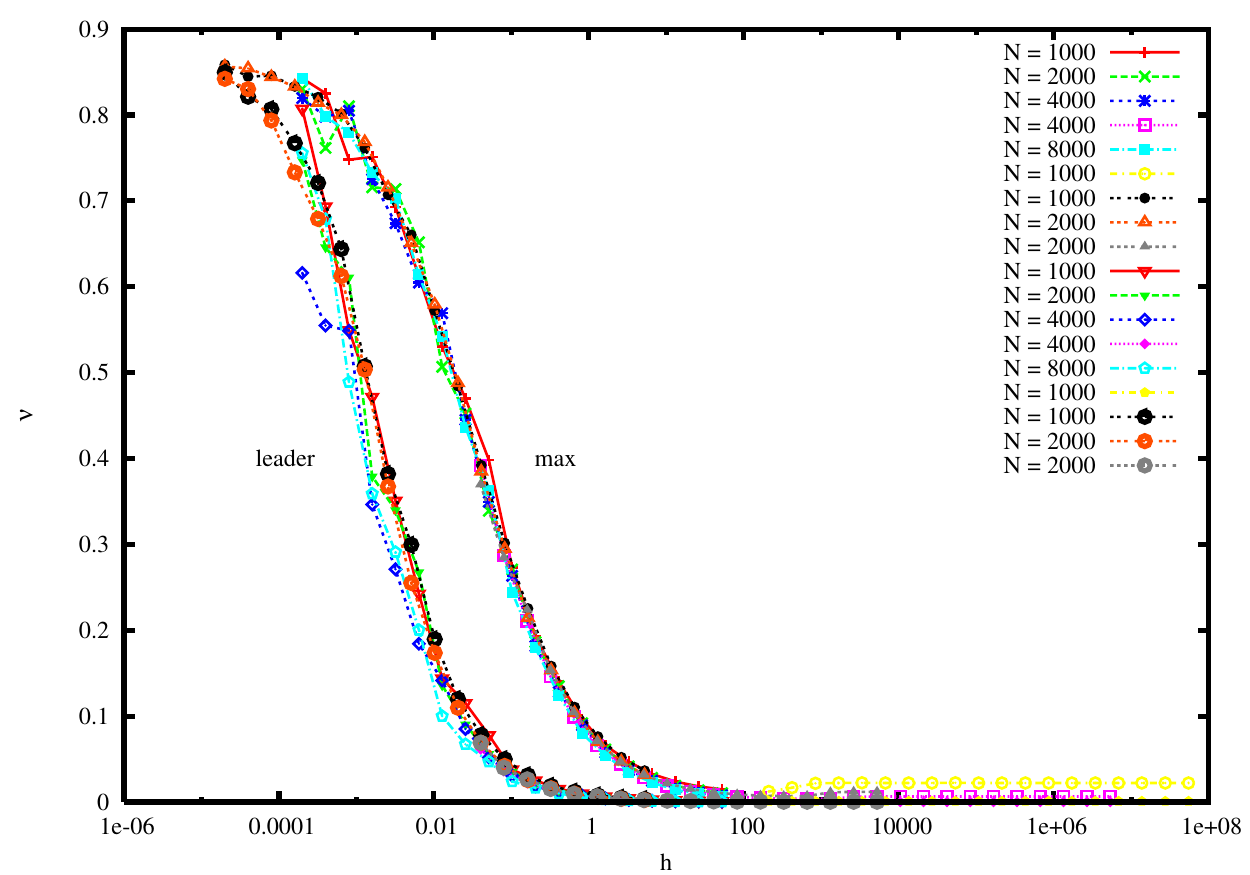}
\caption{Occupation of leader (upper) and max (lower) as a function of $h$ for
  $\mu=0.1$ and various values of $N=1000,2000,4000,8000$. } 
\label{fig6}
\end{center}
\end{figure}

\begin{figure}[htb]
\begin{center}
\includegraphics[width= 0.7 \textwidth]{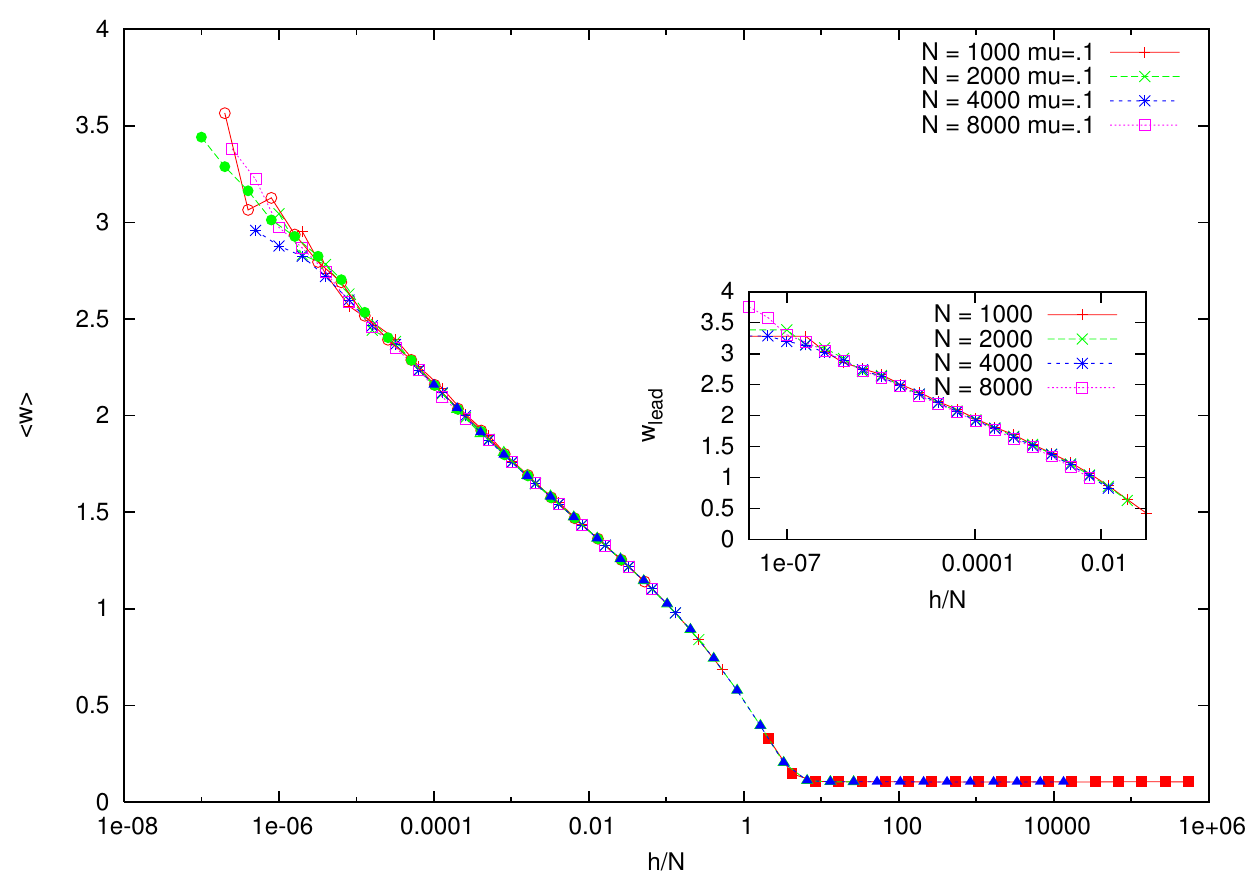}
\caption{Average (main panel) and leader (inset)  fitness as a function of $h/N$ for $\mu=0.1$ and various values of $N=1000,2000,4000,8000$. }
\label{fig7}
\end{center}
\end{figure}

Let us now study the effect of the fitness decay and set $h>0$. We
continue to dwell mainly on the case of lognormal fitness distribution
$\rho(w)$ with parameter $a=0.3$.  In order to study the possibility
of a non zero condensate we observe that while for $h=0$ the maximally
occupied genome should usually coincide with the one of highest
fitness, this is not necessarily true if $h>0$. Let us define
therefore, at any given time, the
\emph{leader} strain in the population as the one that is
populated by the largest fraction of individuals, and the \emph{max}
as the one whose fitness is largest.  In figure~\ref{fig2}
we plot the occupation fraction $\nu$ of the leader as a function of the
mutation rate, for several values of $h$, in a population of size
$N=1000$. The figure shows that while the leader occupation fraction
is a decreasing function of $h$, the error threshold is not destroyed
by the immunity response and the critical value $\mu_\C$ is roughly 
independent of $h$. 
We also observe that a non zero $h$ introduces a natural time scale
into the system, and one should expect that stationarity is reached
after a finite relaxation time, so that the critical value of $\mu$
becomes time independent.  
In figure~\ref{fig3} we compare the fitness of the leader as a
function of time in the condensed phase for $h=0$ to the one
corresponding to a small value of $h$. One can manifestly see
that while stationarity does not hold in the former case, it holds in
the latter. 
In figure~\ref{fig4} we plot the
occupation of the leader as a function of the mutation rate for
various values of $N$. One can see that the error threshold slowly moves to
higher values of $\mu$ as $N$ increases.  
This also appears in the behavior of the leader fitness
$w_{\mathrm{lead}}$ and of the maximal fitness $w_{\mathrm{max}}$
in the system which exhibit a maximum at the error threshold. 
We plot these quantities together with the average fitness in figure~\ref{fig5}. In order to obtain a less cluttered plot, the fitnesses are rescaled by the factor $f(N)=\exp\left(\sqrt{2a^{2}\log(N/\sqrt{2\pi a^{2}})}\right)$, where $a=0.3$ is the width of the distribution of $\log w_{0}$, expected on the basis of extreme-value statistics for the lognormal distribution. The range of variation of this factor is too small to warrant drawing any conclusion from this observation. Notice that while the leader and the maximal fitnesses
reach the maximum at the error threshold, the average fitness
has a maximum at a lower value of $\mu$.  
Close to the maximum of the average fitness the leader and the max
occupation fractions turn out to be roughly independent of $N$, as it
can be seen from figure~\ref{fig6}, which plots these two quantities as
a function of $h$. Conversely, figure~\ref{fig7} shows that the
fitness of the leader and of the max 
are functions of $h/N$: for larger $N$ a proportionally larger $h$ is
needed to decrease the fitness of the same amount. 

\section{Dynamics}
\label{dyn}
\begin{figure}[ht]
\begin{center}
\includegraphics[width= 0.7 \textwidth]{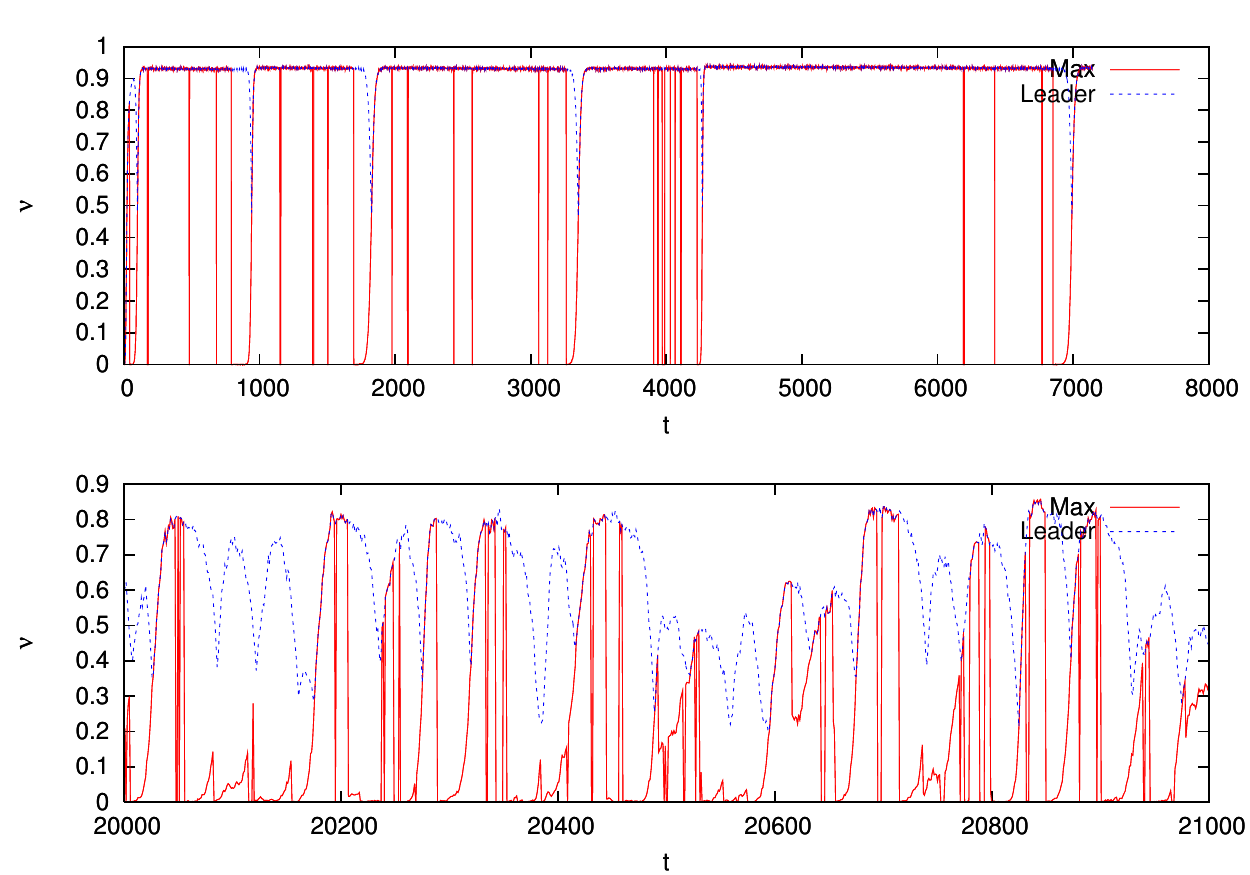}
\caption{Time dependence of the occupation fraction of the max and of
  the leader in the evolving quasispecies regime. Upper panel:  the system is far from
  the threshold, fluctuations are small, the substitution time is the
  only random quantity, $N=40.000$, $\mu=0.05$, $h=0.01$. Lower panel: the
  system is close to the error threshold, fluctuations are large,
  $N=1000$, $\mu=0.1$, $h=0.01$. } 
\label{fig8}
\end{center}
\end{figure}

\begin{figure}[ht]
\begin{center}
\includegraphics[width= 0.7 \textwidth]{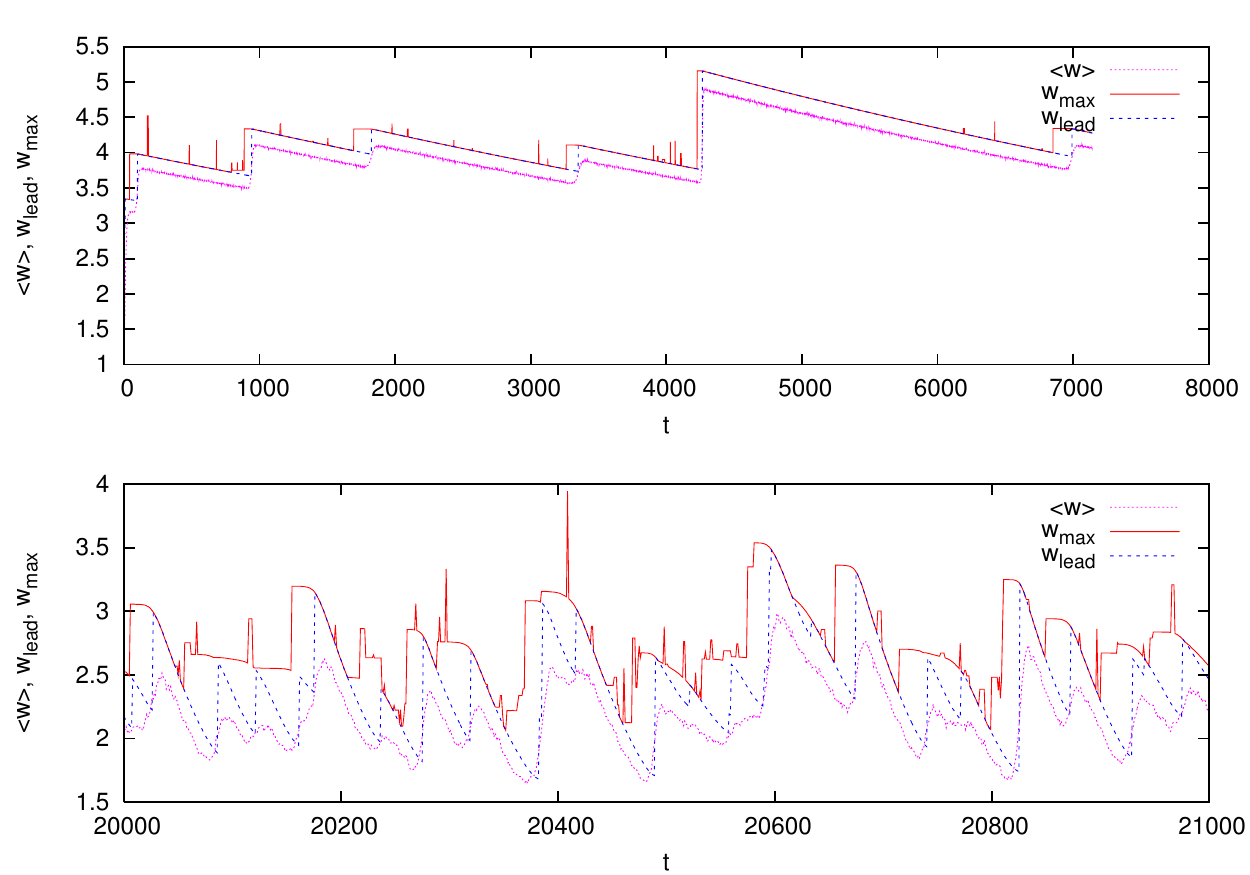}
\caption{Behavior of the fitness in the quasispecies regime. Green:
  max; red: leader. For both panels the parameters are the same as in
  the corresponding ones of figure~\ref{fig8}. } 
\label{fig9}
\end{center}
\end{figure}

\begin{figure}[ht]
\begin{center}
\includegraphics[width= 0.7 \textwidth]{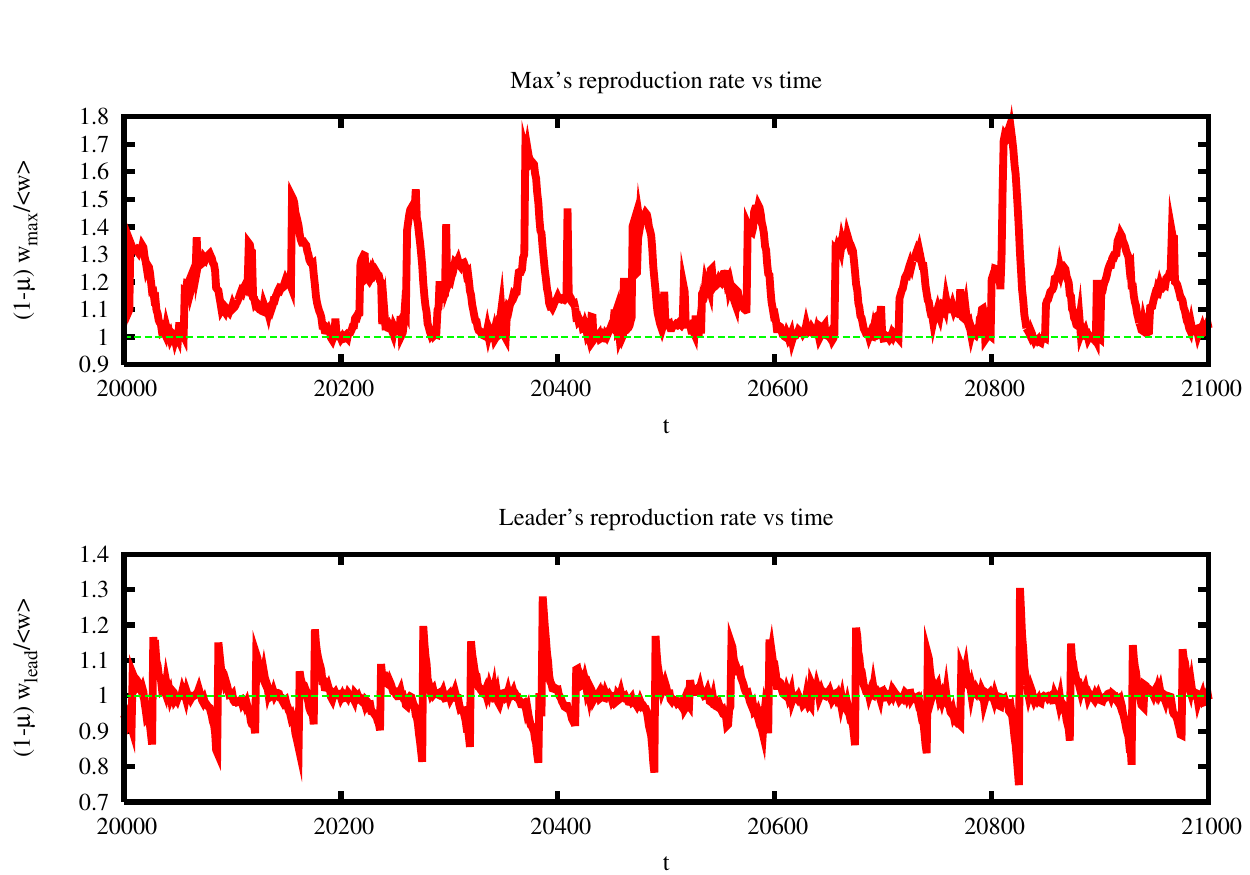}
\caption{Reproduction rate of the max and of the leader, for
  parameters corresponding to the lower panels of \ref{fig8} and
  \ref{fig9}. Notice that the reproduction rate of the max is always
  larger than 1, while the reproduction rate of the leader is
  noticeably different from one only close to a substitution event.} 
\label{fig10.1}
\end{center}
\end{figure}

\begin{figure}[ht]
\begin{center}
\includegraphics[width= 0.7 \textwidth]{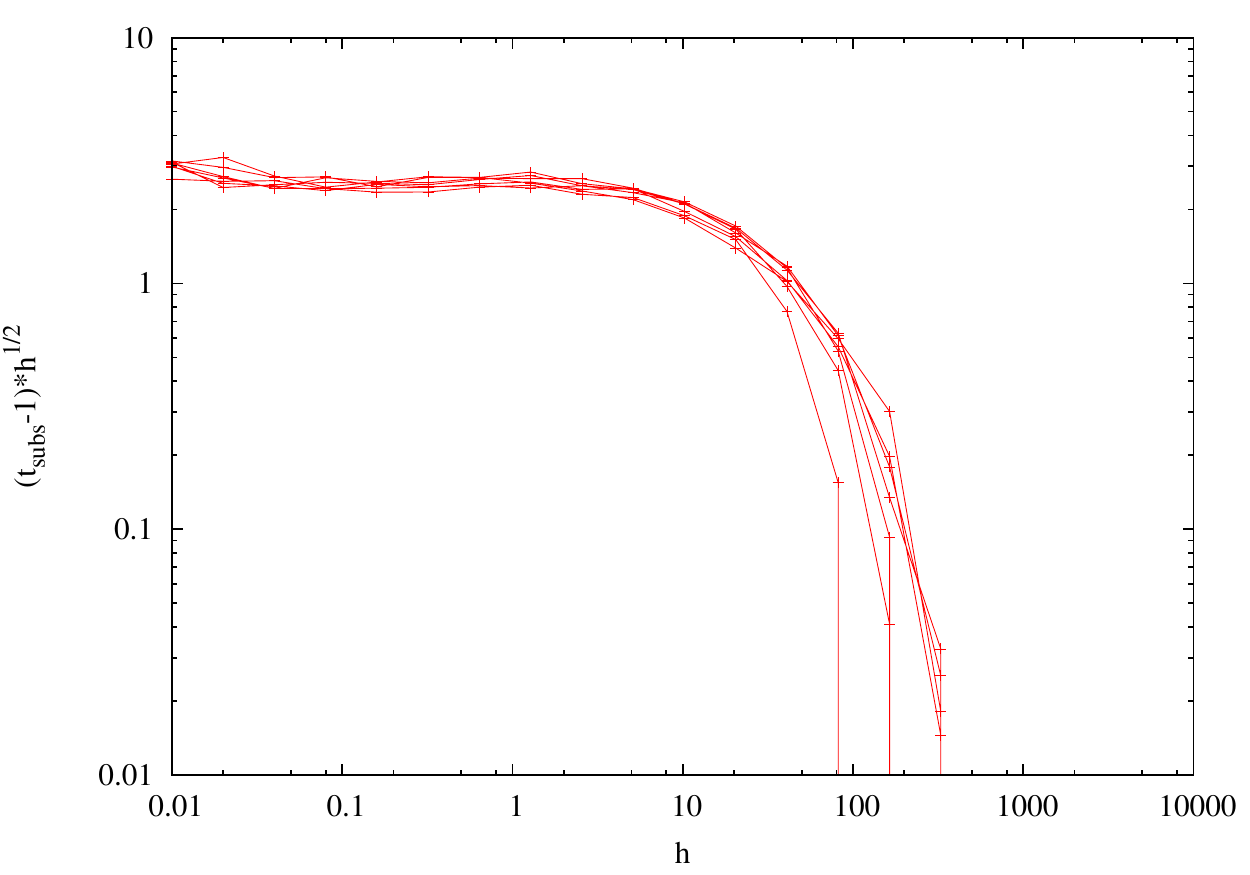}
\caption{Plot of $(t_{\mathrm{subs}}-1) h^{1/2}$ as a function
  of $h$ for $\mu=0.05,0.1,0.15,0.2,0.25,0.3,0.35$. The curves are
  roughly constant in the quasispecies regime.}  
\label{fig10}
\end{center}
\end{figure}

The results of the previous section clearly show that the error
threshold persists even in the presence of a small amount
of immune response in the host. 
In order to gain a better understanding of the evolving quasispecies
regime characterizing the well adapted phase, let us look at the temporal
behavior of the system.  In figure~\ref{fig8} we show the occupation fractions
of the max and of the leader as a function of time, in the well adapted
phase with $\mu=0.1$ and $h=0.01$ in a population of $N=1000$
individuals. We can see that both quantities exhibit a regular behavior
in time, which can be qualitatively described as follows. When a new
fitter mutant 
establishes in the population, it substitutes the previous leader and
remains leader for a while before being at his turn substituted. We
find therefore a kind of punctuated equilibrium dynamics, as announced
in the introduction. 
Figure~\ref{fig9} shows the corresponding behavior of the average
fitness of the leader and of the max respectively. One
sees that the fitness of  
the max stays constant for a while after the emergence of the mutant,
and then, when the max replaces the leader, it rapidly decreases until
a new leader takes over. 
It is interesting to look at the reproduction rate of the leader
strain. As shown in figure~\ref{fig10.1}, this is observed to differ substantially from one only close to
substitution events. Just before substitution, the challenged leader
has a reproduction rate which is smaller than one, while  it is larger
then one just after
substitution. In the periods when the leader largely dominates the
population (dominance periods) the
reproduction ratio is very close to one, and diversity in the
population is maintained by mutants.  
We measured the average duration $t_{\mathrm{subs}}$ of the leadership, i.e., the time
interval between successive substitutions. In figure~\ref{fig10} we plot
$(t_{\mathrm{subs}}-1) h^{1/2}$ vs.~$h$, showing that this quantity it is roughly proportional to $h^{-1/2}$, and essentially independent of $\mu$, in the quasispecies regime. 
Given the natural decay time of fitness on time scales of order $1/h$ it
 would have been natural to hypothesize a substitution time of the same order. 
Our numerical result show that substitutions being separated by times of order $h^{-1/2}$ 
are much more frequent.   

\section{The Diluted Champion Process} 
\label{dcp}
Certain aspects of the evolving quasispecies dynamics at low $h$ and
$\mu$ can be interpreted by introducing an effective stochastic process
for the substitution. We start by defining the Champion Process where
at each time there is a leader. The performance of the leader,
which is represented by a fitness value, declines exponentially in
time: $W_{t+1}=\lambda W_t$ with $\lambda<1$. The leader is
challenged at regular intervals of time, and the challenger's fitness
is extracted randomly from a distribution $\rho^*(W)$. If the fitness
of the challenger exceeds the one of the leader in charge, the
challenger substitutes the old leader. Otherwise there is no
substitution and the old champion remains in charge.  The Diluted
Champion Process (DCP) is a simple variation of the process where the
substitution process is not deterministic. Substitutions can occur
with a certain probability that depends on the fitness ratio between
the challenger and the champion in charge.  This model can obviously 
be adapted to competition situations where the performances of the
individuals naturally decline in time.  

Now, in the evolving quasispecies regime of our model, while 
the substitution times and the leaders fitnesses are random quantities,
the dynamics between punctuations is essentially deterministic. 
Moreover, the champion substitution process requires times that are
much shorter than the typical times of change of the fitness. Indeed as
figures~\ref{fig8} and~\ref{fig9} show, the fitness of the champion
starts to decay as soon as the new challenger takes over.  
These observations make the DCP description pertinent for our model.
During the dominance periods, at each time step the champion is
challenged by a number of mutants $M$ Poisson distributed with  
$\langle M\rangle=N\mu$, with random fitnesses drawn from $\rho(w)$. 
By elementary extreme values statistics, the best fit challenger has
therefore a fitness distribution given by
$\rho^*(w)=\mu N \rho(w)\, \rme^{-\mu N \int_w^\infty dw'\;
  \rho(w')}$. This will have a fixation probability $f(w/w_\leader)$,  
which can be estimated by the Kimura expression:
$f(x)=(1-x^{-1})/(1-x^{-N})$.  
Thus the probability of survival in one round of a
leader with fitness $w_\leader$ is given by 
\begin{eqnarray}
\Phi(w_\leader) 
=1-\int \rmd w\; \rho^* (w)f\left(\frac{w}{w_\leader}\right).
\end{eqnarray}
At this point we will assume, consistently with the simulations, that
for small $h$ we can neglect the fixation  time with respect to the
leader strain lifetime, and suppose that for 
a champion $w_{t+1}=\lambda w_t$, with a constant $\lambda$ equal to  
$\lambda=\exp(-n^* h)$ where $n^*$ is the condensate fraction.  In the case of the DCP, one can safely assume $n^{*}\simeq 1$, leading to $\lambda = \E^{-h}$.

\begin{figure}[ht]
\begin{center}
\includegraphics[width= 0.7 \textwidth]{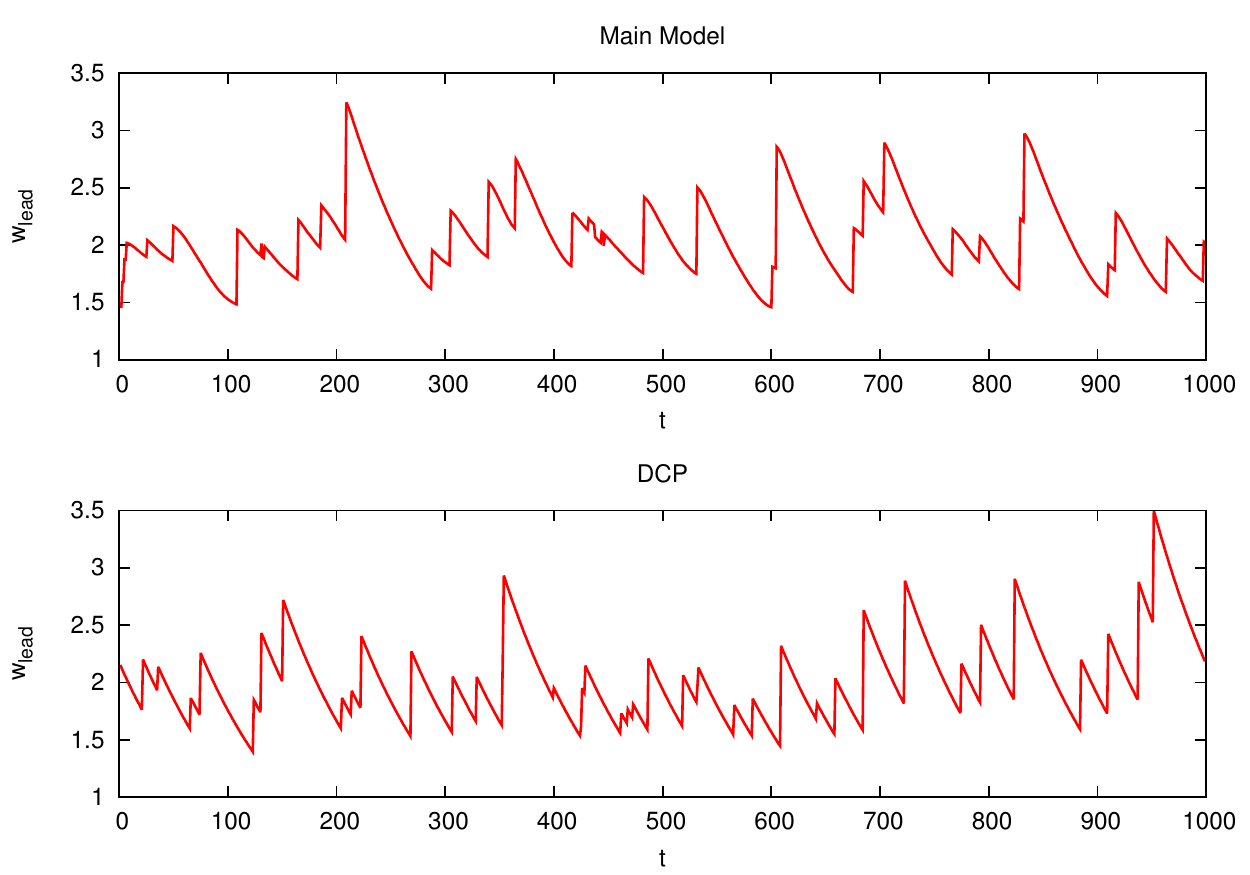}
\caption{Leader's fitness  as a function of time for the main model
  and for the DCP with $N=400$,
  $\mu=0.05$, $h=0.01$ and $a=0.3$. The DCP clearly reproduces the qualitative features of the main model.} 
\label{champ}
\end{center}
\end{figure}

It is possible to formally write  the basic formula for the
probability that a champion  
with initial fitness $w_0$ is substituted after exactly $t+1$ time steps 
by a new champion with fitness $w_1$, namely:
\begin{eqnarray}
P_t(w_1\given w_0)=
\prod_{s=1}^t \Phi(w_0  \lambda^s ) \rho^*(w_1)
f\left(\frac{w_1}{w_0\lambda^t}\right). 
\end{eqnarray}
From this equation all the the properties of the DCP can be in
principle derived.   
One can express the conditional probability that the leader fitness equals $w_1$
given the previous leader fitness $w_0$ by
\begin{eqnarray}
M(w_1\given w_0)=\sum_{t=0}^\infty P_t(w_1\given w_0).
\end{eqnarray}
Then one can in principle obtain the stationary distribution of the
leader fitness $\mu(w)$ from 
\begin{eqnarray}
\mu(w_1)=\int \rmd w_0\; M(w_1\given w_0)\mu(w_0),
\end{eqnarray}
and the distribution of the substitution time at stationarity from
\begin{eqnarray}
q(t)=
\int \rmd w_0\; dw_1\;  P_t(w_1\given w_0)\mu(w_0).
\end{eqnarray}
Unfortunately these equations are not easily analyzed, so in order to
test the validity of this simple effective model we had to simulate
it. In figure~\ref{champ} we show the typical evolution of the
leader's fitness as a function of time.  The DCP clearly reproduces the qualitative features of our model.

\section{Conclusions and perspectives}
We have introduced a simple effective model describing the evolution of a
population of pathogens in the presence of the immunological response
of the host. The model is simple enough to be amenable to a systematic
numerical investigation, and allows one to identify a parameter
region in which the coexistence of a well-defined
quasispecies and of a fast turnover of the dominant strain is
be quite robust. The behavior of the model in this region can be
understood in terms of a further simplified model: the Diluted Champion 
Process. Our model can be interpreted in several ways: on the
one hand, as describing the evolving viral strains present in the host
population, 
e.g., in the case of the annual influenza epidemics; on the other
hand, as the evolution of the viral population in a chronic infection
of a single host individual, as in intra-host HIV evolution.
In this case, fluctuations in the strength of immune response could
cause proliferation of the number of strains and then the 
failure of the immunity system to downregulate all the viral strains 
(see, e.g., ~\cite{Nowak}). It is possible to obtain a more realistic
description of this regime by, e.g., relaxing the fixed population
constraint. It is also possible to consider a higher degree of
correlation in the fitness landscape, in order to understand the
extent of the stability of the punctuated equilibrium regime.

We are working on a generalization of our model taking into account
the existence of different regions. They can be understood as
different climatic regions in the case of epidemics, or localization
in different organs in the case of intra-host infection. This
investigation will allow us to understand better the different roles
played by different regions: seeder,
reservoirs, etc., as observed in recent analyses of epidemiological
data collected 
in the last years~(\cite{Rambaut,Russell}), or to identify possible
gene-surfing phenomena in the diffusion of new strains from external
regions~(\cite{Hallatschek_Nelson}).

\ack LP acknowledges the support of FARO.

\section*{References}
\bibliography{flu2}

\end{document}